\documentstyle[twoside,fleqn,epsf,espcrc2]{article}
\input psfig

\newcommand{\AmS}{{\protect\the\textfont2
  A\kern-.1667em\lower.5ex\hbox{M}\kern-.125emS}}

\title{Beyond The Standard Model:  This Time for Real}

\author{Frank Wilczek\address{Institute for Advanced Study,\\ School
        of Natural Sciences, Olden Lane, \\ Princeton, New Jersey
        08540} \thanks{Invited Theoretical Summary Talk presented at
        XVIII International Conference on Neutrino Physics and
        Astrophysics, Takayama, Japan, June 4-9, 1998.  Research
        supported in part by DOE grant DE-FG02-90ER40542.
        IASSNS-HEP98/79}}

\begin{document}

\begin{abstract}
The value of the neutrino mass reported by the SuperK collaboration
fits beautifully into the framework of gauge theory unification.  Here
I justify this claim, and review the other main reasons to believe in
that framework.  Supersymmetry and $SO(10)$ symmetry are important
ingredients; nucleon instability is a dramatic consequence.

\end{abstract}

\maketitle

It has been a great privilege to attend this conference, which I am
sure the future will regard as historic.  I want to thank the
organizers for making it in every way a very enjoyable experience, as
well.

Undoubtedly it will take us, collectively, many years to do full
justice to the wonderful discovery announced here, that neutrinos have
non-zero mass.  Many important tasks remain at the level of pure
phenomenology, most obviously perhaps that of integrating the firm
atmospheric oscillation results with the long-standing but still
confusing solar neutrino anomalies, and the possible hints from LSND
of a third distinct effect.  However I am going to indulge myself by
leaping over these vital issues, to discourse and speculate on the
larger implications of the discovery for fundamental physics.  Some of
us have been hoping for many years to see results of this kind.  Now
that they are coming in, we look forward with both eagerness and
trepidation to the confrontation of our dreams with reality.  Let me
remind you what's at stake.

\section{A New Scale}

It is important to realize that the degrees of freedom of
the Standard Model permit neutrino masses.  A minimal implementation
of the construction requires an interaction of the type 

\begin{equation}
\Delta
{\cal L} ~=~ \kappa_{ij} L^{\alpha a i} L^{\beta b j} \epsilon_{\alpha
\beta} \phi^\dagger_a \phi^\dagger_b + {\rm h. c.} ~, 
\label{eq:newL}
\end{equation}
where $i$ and $j$ are family indices; $\kappa_{ij}$ is a symmetric
matrix of coupling constants; the $L$ fields are the left-handed
doublets of leptons, with Greek spinor indices, early Roman weak
$SU(2)$ indices, and middle Roman flavor indices; and finally $\phi$
is the Higgs doublet, with its weak $SU(2)$ index.  Two-component
notation has been used for the spinors, to emphasize that this way of
forming mass terms, although different from what we are used to for
quarks and charged leptons, is in some sense more elementary
mathematically.  $\Delta {\cal L}$ becomes a neutrino mass term when
the $\phi$ field is replaced by its vacuum expectation value $\langle
\phi^a \rangle = v \delta^a_1$.

Although this Eq. (\ref{eq:newL}) is a possible interaction for the degrees of
freedom in the Standard Model, it is usually considered ``beyond''
the Standard Model, for a very good reason.  The new term differs from
the terms traditionally included in the Standard Model in that the
product of fields has mass dimension 5, so that the coefficient
$\kappa$ must have mass dimension -1.  In the context of quantum field
theory, it is a nonrenormalizable interaction.  When one includes it
in virtual particle loops, one will find amplitudes containing the
dimensionless factors of the type $\kappa \Lambda$, where $\Lambda$ is
an ultraviolet cutoff.  In this framework, therefore, one cannot
accept $\Delta {\cal L}$ as an elementary interaction.  It can only be
understood within a larger theoretical context.

Given a numerical value for the neutrino mass, we can infer a scale
beyond which $\Delta {\cal L}$ cannot be accurate, and degrees of
freedom beyond the Standard Model must open up.  To get oriented, let
us momentarily pretend that $\kappa$ is simply a number instead of a
matrix, and that $m = 10^{-2}$ eV is the neutrino mass.  Then, using
$v = 250$ GeV for the vacuum expectation value, we calculate 

\begin{equation}
1/M \equiv \kappa = m/v^2 = 1/(6 \times 10^{15}~ {\rm GeV}) ~~.
\label{eq:newScale}
\end{equation}

When energy and momenta of order $M$ begin to circulate in loops the
form of the interaction must be modified.  Otherwise the dangerous
factor $\kappa \Lambda$ will become larger than unity, inducing large
and uncontrolled radiative corrections to all processes, and rendering
the success of the Standard Model accidental.

Thus we trace the ``absurdly small'' value of the observed
neutrino mass scale to an ``absurdly large'' fundamental mass
scale.  As I shall now discuss, this new scale provides, on the face
of it, a wonderful confirmation of our best developed ideas for
unification beyond the Standard Model.

Of course, experts will recognize that the foregoing argument is
oversimplified; in due course, I shall revisit it in a more critical
spirit.

\section{Two Pillars of Unification }

\begin{figure}[htb]
\centerline{\psfig{figure=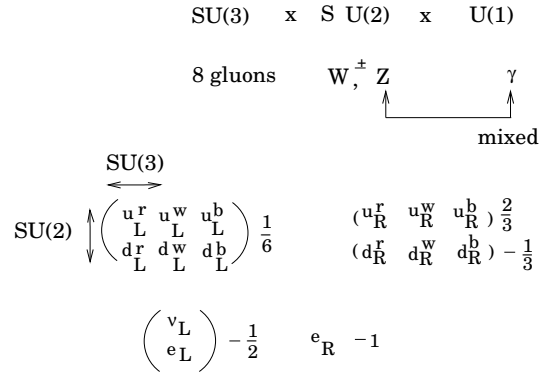,width=7cm,angle=-90}}
\vglue-.4in
\caption[]{The gauge groups of the standard model, and the
fermion multiplets with their hypercharges.}
\label{fig:su1.ps}
\end{figure}

The standard model of particle physics is based upon the
gauge groups SU(3)$\times$SU(2)$\times$U(1) of strong, electromagnetic
and weak interactions acting on the quark and lepton multiplets
as shown in Figure \ref{fig:su1.ps}.

In this Figure I have depicted only one family (u,d,e,$\nu_e$) of
quarks and leptons; in reality there seem to be three families which
are mere copies of one another as far as their interactions with the
gauge bosons are concerned, but differ in mass.  Actually in the
Figure I have ignored masses altogether, and allowed myself the
convenient fiction of pretending that the quarks and leptons have a
definite chirality -- right- or left-handed -- as they would if they
were massless.  The more precise statement, of course, is that the
gauge bosons couple to currents of definite chirality.  The chirality
is indicated by a subscript R or L.  Finally the little number beside
each multiplet is its assignment under the U(1) of hypercharge, which
is the average of the electric charge of the multiplet.

While little doubt can remain that the Standard Model is essentially
correct, a glance at Figure \ref{fig:su1.ps} is enough to reveal that it is not a
complete or final theory.  To remove its imperfections, while building
upon its solid success, is a worthy challenge.

There are two improvements on the Standard Model that are so deeply
suggested in its structure, that I think it is perverse to deny them.
Let me briefly recall these two pillars of unification:

\section{Gauge Group and Fermion Unification} 

Given that the strong interactions are governed by transformations
among three colors, and the weak by transformations between two
others, what could be more natural than to embed both theories into a
larger theory of transformations among all five colors \cite{Pati-Salam}?

This idea has the additional attraction that an extra U(1) symmetry
commuting with the strong SU(3) and weak SU(2) symmetries
automatically appears, which we can attempt to identify with the
remaining gauge symmetry of the standard model, that is hypercharge.
For while in the separate SU(3) and SU(2) theories we must throw out
the two gauge bosons which couple respectively to the color
combinations R+W+B and G+P, in the SU(5) theory we only project out
R+W+B+G+P, while the orthogonal combination (R+W+B)-${3\over 2}$(G+P)
remains.

Finally, the possibility of unified gauge symmetry breaking is
plausible by analogy; after all, we know for sure that gauge symmetry
breaking occurs in the electroweak sector.

Georgi and Glashow \cite{gegl} showed how these ideas can be used to
bring some order to the quark and lepton sector, and in particular to
supply a satisfying explanation of the weird hypercharge assignments
in the standard model.  As shown in Figure \ref{fig:colors}, the five scattered
SU(3)$\times$SU(2)$\times$U(1) multiplets get organized into just two
representations of SU(5).

In making this unification it is necessary to allow transformations
between (what were previously considered to be) particles and
antiparticles of the same chirality, and also between quarks and
leptons.  It is convenient to work with left-handed fields only.
Since the conjugate of a right-handed field is left-handed, we don't
lose anything by doing so -- though we must shed traditional
prejudices about a rigorous distinction between matter and antimatter,
since these get mixed up.  Specifically, it will not be possible to
declare that matter is what carries positive baryon and lepton number,
since the unified theory does not conserve these quantum numbers.

As shown in Figure \ref{fig:colors}, there is one group of ten left-handed fermions
that have all possible combinations of one unit of each of two
different colors, and another group of five left-handed fermions that
each carry just one negative unit of some color.  These are the
ten-dimensional antisymmetric tensor and the complex conjugate of the
five-dimensional vector representation, commonly referred to as the
``five-bar''.  In this way, {\it the structure of the standard model,
with the particle assignments gleaned from decades of experimental
effort and theoretical interpretation, is perfectly reproduced by a
simple abstract set of rules for manipulating symmetrical symbols}.
Thus for example the object RB in this Figure has just the strong,
electromagnetic, and weak interactions we expect of the complex
conjugate of the right-handed up-quark, without our having to instruct
the theory further.

A most impressive, though simple, exercise is to work out the
hypercharges of the objects in Figure \ref{fig:colors} and checking against what you
need in the Standard Model.  These ugly ducklings of the Standard
Model have matured into quite lovely swans.

\bigskip

\begin{figure}
\parskip=0pt{
\underline{SU(5):  5 colors RWBGP}

$\underline{10}$: 2 different color labels (antisymmetric tensor)

$$\matrix{\rm u_L:&\rm RP,&\rm WP,&\rm BP\cr
\rm d_L:&\rm RG,&\rm WG,&\rm BG\cr
\rm u{^c_L}:&\rm RW,&\rm WB,&\rm BR\cr
&\rm (\bar B)&\rm (\bar R)&\rm (\bar W)\cr
\rm e{^c_L}:&\rm GP&&\cr
&(\ )&&\cr
}
\pmatrix{0&\rm u^c&\rm u^c&\rm u&\rm d\cr
&0&\rm u^c&\rm u&\rm d\cr
&&0&\rm u&\rm d\cr
&*&&0&\rm e\cr
&&&&0\cr}$$

$\underline{\bar 5}$: 1 anticolor label

$$\matrix{\rm d{^c_L}:&\rm \bar R,&\rm  \bar W,&\rm  \bar B\cr
\rm e_L:&\rm \bar P&&\cr
\nu_{\rm L}:&\rm \bar G&&\cr
}
\matrix{\rm (d^c&\rm d^c&\rm d^c&{\rm e}&\nu)\cr}$$
\def\boxtext#1{%
\vbox{%
\hrule
\hbox{\strut \vrule{} #1 \vrule}%
\hrule
}%
}
\centerline{
\vbox{\offinterlineskip
\hbox{\boxtext{\rm Y $= -{1\over 3}$ (R+W+B) $+{1\over 2}$ (G+P)}}
}}
}
\vglue-.4in
\caption[]{Unification of fermions in SU(5)\label{fig:colors}}
\end{figure}

\section{Coupling Constant Unification}

We have just seen that simple unification schemes are spectacularly
successful at the level of classification.  New questions arise when
we consider dynamics.

Part of the power of gauge symmetry is that it fully dictates the
interactions of the gauge bosons, once an overall coupling constant
is specified.  Thus if SU(5) or some higher symmetry were exact, then
the fundamental
strengths of the different color-changing interactions would have
to be equal, as would the
(properly normalized) hypercharge coupling strength.  In reality the
coupling strengths of the gauge bosons in SU(3)$\times$SU(2)$\times$U(1)
are not observed to be equal, but rather follow the pattern
$g_3 \gg g_2 > g_1$.

Fortunately, experience with QCD emphasizes that couplings ``run''\cite{dgfw}.
The physical mechanism of this effect is that in quantum field theory
the vacuum must be regarded as a polarizable medium, since virtual
particle-anti-particle pairs can screen charge.  For charged gauge
bosons, as arise in non-abelian theories, the paramagnetic
(antiscreening) effect of their spin-spin interaction dominates, which
leads to asymptotic freedom.  As Georgi, Quinn, and Weinberg pointed
out \cite{gqw}, if a gauge symmetry such as SU(5) is spontaneously broken
at some very short distance then we should not expect that the
effective couplings probed at much larger distances, such as are
actually measured at practical accelerators, will be equal.  Rather
they will all have have been affected to a greater or lesser extent by
vacuum screening and anti-screening, starting from a common value at
the unification scale but then diverging from one another.  The
pattern $g_3 \gg g_2 > g_1$ is just what one should expect, since the
antiscreening effect of gauge bosons is more pronounced for larger
gauge groups.

\begin{figure}[!t]
\centerline{\psfig{figure=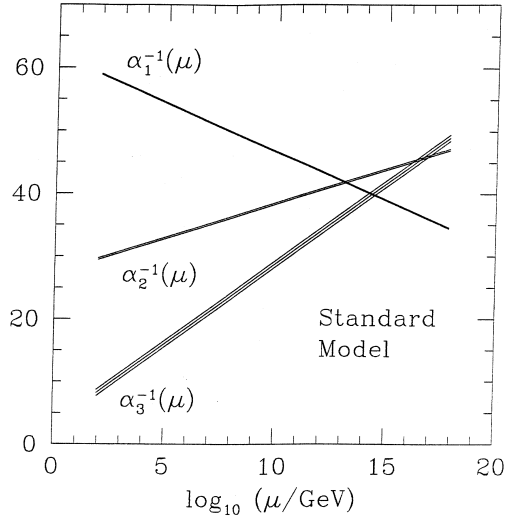,width=7cm}}
\vglue-.4in
\caption[]{The failure of the running couplings, normalized according
to SU(5) and extrapolated taking into account only the virtual
exchange of the ``known'' particles of the standard model (including
the top quark and Higgs boson) to meet.  Note that only with fairly
recent experiments \cite{EWG}, which greatly improved the precision
of the determination of low-energy couplings, has the discrepancy
become significant.}
\label{fig:cosmo6.EPS}
\end{figure}

The running of the couplings gives us a truly quantitative handle on
the ideas of unification.  To specify the relevant aspects of
unification, one basically needs only to fix two parameters: the scale
at which the couplings unite, (which is essentially the scale at which
the unified symmetry breaks), and their common value when they unite.
Given these, one calculates three outputs, the three {\it a priori\/}
independent couplings for the gauge groups in
SU(3)$\times$SU(2)$\times$U(1).  Thus the framework is eminently
falsifiable.  The astonishing thing is, how close it comes to working
(Figure \ref{fig:cosmo6.EPS}).


The GQW calculation is remarkably successful in
explaining the observed hierarchy $g_3 \gg g_2 > g_1$ of
couplings and the approximate stability of the proton.
In performing it, we assumed that the known and
confidently expected particles of the standard model exhaust
the spectrum up to the unification scale, and that the
rules of quantum field
theory could be extrapolated without alteration
up to this mass scale -- thirteen orders
of magnitude beyond the domain they were designed to describe.
It is a triumph for minimalism, both existential and conceptual.

On closer inspection, however, 
it is not quite good enough.  Accurate modern measurements
of the couplings show a small but definite discrepancy between the
couplings, as appears in Figure \ref{fig:cosmo6.EPS}.  And heroic dedicated experiments to
search for proton decay did not find it \cite{blewitt}; they currently
exclude the minimal SU(5) prediction
$\tau_p \sim 10^{31}~{\rm yrs.}$ by about two orders of magnitude.

If we just add particles in some haphazard
way things will
only get
worse: minimal SU(5) nearly works, so a generic perturbation
will be deleterious.  Even if some {\it ad hoc\/}
prescription could be made to work,
that would be a disappointing outcome from what
appeared to be one of our most precious, elegantly
straightforward clues regarding physics well
beyond the Standard Model.

\begin{figure}[!t]
\centerline{\psfig{figure=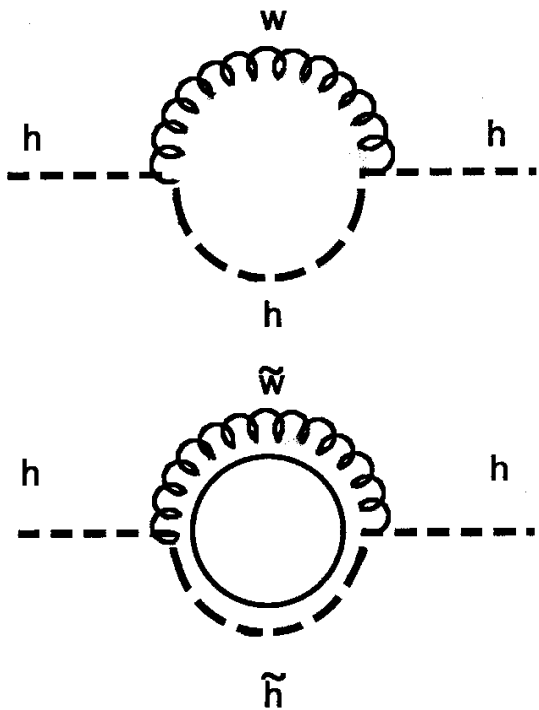,width=7cm}}
\caption[]{Contributions to the Higgs field self-energy.  These graphs
give contributions to the Higgs field self-energy which separately are
formally quadratically divergent, but when both are included the
divergence is removed.  In models with broken supersymmetry a finite
residual piece remains.  If one is to obtain an adequately small
finite contribution to the self-energy, the mass difference between
Standard Model particles and their superpartners cannot be too great.
This -- and essentially only this -- motivates the inclusion of
virtual superpartner contributions in Figure \ref{fig:cosfig7.eps} beginning
at relatively low scales.}
\label{fig:cosfig8.eps}
\end{figure}

Fortunately, there is a compelling escape from this impasse.  That is
the idea of supersymmetry \cite{ferra}.  Supersymmetry is certainly not a
symmetry in nature: for example, there is certainly no bosonic
particle with the mass and charge of the electron.  However there are
several reasons for thinking that supersymmetry might be
spontaneously, and only relatively mildly broken, so that the
superpartners are no more massive than $\approx$ 1 Tev.  The most
concrete arises in calculating radiative corrections to the (mass)$^2$
of the Higgs particle from diagrams of the type shown in 
{}Figure \ref{fig:cosfig8.eps}.
One finds that they make an infinite, and also large, contribution.
By this I mean that the divergence is quadratic in the ultraviolet
cutoff.  No ordinary symmetry will make its coefficient vanish.  If we
imagine that the unification scale provides the cutoff, we will find,
generically, that the radiative correction to the (mass)$^2$ is much
larger than the total value we need to match experiment.  This is an
ugly situation.

In a supersymmetric theory, if the supersymmetry is not too badly
broken, it is possible to do better.  For any set of virtual particles
that might circulate in the loop there will be another graph with
their supersymmetric partners circulating.  If the partners were
accurately degenerate, the contributions would cancel.  Taking
supersymmetry breaking into account, the threatened quadratic
divergence will be cut off only at virtual momenta such that the
difference in (mass)$^2$ between the virtual particle and its
supersymmetric partner is negligible.  Notice that we will be assured
adequate cancellation if and only if supersymmetric partners are not
too far split in mass -- in the present context, if the splitting
times the square root of the fine structure constant is not much
greater than the weak scale.

The effect of low-energy supersymmetry on the running of the couplings
was first considered long ago \cite{DRW}, in advance of the precise
measurements of low-energy couplings or of the modern limits on
nucleon decay.  One might have feared that such a huge expansion of
the theory, which essentially doubles the spectrum, would utterly
destroy the approximate success of the minimal SU(5) calculation.
This is not true, however.  To a first approximation since
supersymmetry is a space-time rather than an internal symmetry it does
not affect the group-theoretic structure of the calculation.

Thus to a first approximation the absolute
rate at which the couplings run
with momentum is affected, but not the relative rates.  The main effect
is that the supersymmetric partners of the color gluons, the gluinos,
weaken the asymptotic freedom of the strong interaction.  Thus they
tend to
make its effective
coupling decrease and approach the others more slowly.  Thus
their merger requires a longer lever arm,
and the scale at which the couplings meet increases by an order of
magnitude or so, to about 10$^{16}$ Gev.

I want to emphasize that this very large new mass scale has emerged
unforced from the internal logic of the Standard Model itself.

Its value is important in several ways.  First, it explains why the
exchange of gauge bosons that are in SU(5) but not in
SU(3)$\times$SU(2)$\times$U(1) does not lead to catastrophically quick
nucleon decay.  Second, it brings us close to the Planck scale $M_{\rm
Planck} \sim 10^{19}~{\rm Gev}$ at which exchange of gravitons
competes quantitatively with the other interactions.  Because $M_{\rm
un.}$ is significantly smaller than the Planck mass, we need not be
too nervous about the neglect of quantum gravity corrections to our
calculation; but because it is not absurdly smaller, we can feel
encouraged for the prospect of unification including both gravity and
gauge forces.

Finally, as I shall be emphasizing, it can hardly be accidental that
the unification scale found here is so close to the scale we
previously gleaned from the neutrino mass.

\begin{figure}[!t]
\centerline{\psfig{figure=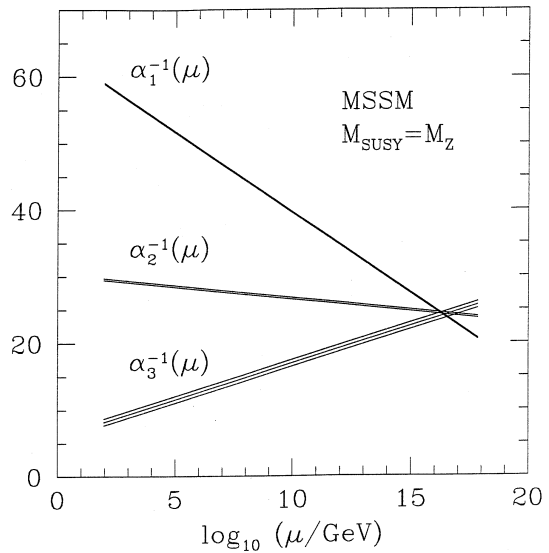,width=7cm}}
\vglue-.4in
\caption[]{When the exchange of the virtual particles necessary
to implement low-energy supersymmetry, a calculation along the lines
of Figure \ref{fig:cosmo6.EPS} comes into adequate agreement with experiment.}
\label{fig:cosfig7.eps}
\end{figure}

There is another effect of low-energy supersymmetry on the running of
the couplings, which although quantitatively small is of prime
interest.  There is an important exception to the general rule that
adding supersymmetric partners does not immediately (at the one loop
level) affect the {\it relative\/} rates at which the couplings run.
That rule works for particles that come in complete SU(5) multiplets,
such as the quarks and leptons, or for the supersymmetric partners of
the gauge bosons, because they just renormalize the existing, dominant
effect of the gauge bosons themselves.  However there is one peculiar
additional contribution, from the Higgs doublets.  It affects only the
weak SU(2) and hypercharge U(1) couplings.  The net affect of doubling
the number of Higgs fields (as, for slightly technical reasons, one
must) and including their supersymmetric partners is a sixfold
enhancement of the Higgs field contribution to the running of weak and
hypercharge couplings.  This causes a small, accurately calculable
change in the unification of couplings calculation.  From Figure \ref{fig:cosfig7.eps}
you see that it is a most welcome one.  Indeed, in the minimal
implementation of supersymmetric unification, it puts the running of
couplings calculation right back on the money \cite{ellis}.

Since the running of the couplings with scale is logarithmic, the
unification of couplings calculation is not terribly sensitive to the
exact scale at which supersymmetry is broken, say between 100 Gev and
10 Tev.  There have been attempts to push the calculation further, in
order to address this question of the supersymmetry breaking scale,
but there are many possibilities, and it is difficult to decide among
them.  An intriguing recent contribution is \cite{smbarr}.

\section{SO(10), and a Third Pillar} 

There is a beautiful extension of SU(5) to
the slightly larger group SO(10).  With this extension, one can unite
all the observed fermions 
of a family, plus one more, into a {\it single\/}  multiplet \cite{georgi}.  
The relevant
representation for the fermions is a 16-dimensional spinor representation.
Some of its features are depicted in Figure \ref{fig:5bit}.

\begin{figure}
\parskip=0pt
\hglue0.75in\underline{SO(10): 5 bit register}
\vglue-.15in
$$(\pm \pm \pm \pm \pm)\ \ :\ \ \underline{\rm even}\ \  \# \  of\  -$$
$$10:\matrix{(++-|+-)&6&\rm (u_L,d_L)\cr
(+--|++)&3&\rm u{^c_L}\cr
(+++|--)&1&\rm e{^c_L}\cr}$$

$$\bar 5:\matrix{(+--|--)&\bar 3&\rm d{^c_L}\cr
(---|+-)&\bar 2&{\rm (e_L},\nu_L)\cr}$$
\nopagebreak
$$1:\matrix{(+++|++)&1&\rm N_R\cr}$$
\vglue-.4in
\caption[]{Unification of fermions in SO(10).  The rule is that all
possible combinations of 5 + and - signs occur, subject to the
constraint that the total number of - signs is even.  The SU(5) gauge
bosons within SO(10) do not change the numbers of signs, and one sees
the SU(5) multiplets emerging.  However there are additional
transformations in SO(10) but not in SU(5), which allow any fermion to
be transformed into any other.\label{fig:5bit}}
\end{figure}

In addition to the conventional quarks and leptons the SO(10) spinor
contains an additional particle, an SU(3)$\times$SU(2)$\times$U(1)
singlet.  (It is even an SU(5) singlet.)  Usually when a theory
predicts unobserved new particles they are an embarrassment.  But
these N particles -- there are three of them, one for each family --
are a notable exception.  Indeed, they are central to the emerging
connection between neutrino masses and unification \cite{gell}.

Because the $N^i$ are singlets, mass terms of the type

\begin{equation}
\Delta {\cal L}_N ~=~ \eta_{ij} N^{\alpha i} N^{\beta j}~ 
\epsilon_{\alpha\beta}
\label{eq:NMassTerm}
\end{equation}
with $\eta_{ij}$ a symmetric coupling matrix, are consistent with
$SU(3)\times SU(2) \times U(1)$ symmetry.  This term of course greatly
resembles the effective interaction responsible for neutrino masses,
Eq. (\ref{eq:newL}), but the difference is conceptually crucial.  Because the Ns are
Standard Model singlets the Higgs doublets that occurred in
Eq. (\ref{eq:newL}) need not appear here.  A consequence is that the
operators appearing in Eq. (\ref{eq:NMassTerm}) have mass dimension 3,
so that the $\eta_{ij}$ must have mass dimension +1.  This interaction
therefore does not bring in any ultraviolet divergence problems.

What sets the scale for $\eta$?  Although Eq. (\ref{eq:NMassTerm}) is
consistent with Standard Model gauge symmetries, or even $SU(5)$, it
is not consistent with $SO(10)$.  Indeed for the product of spinor 16
we have the decomposition 16$\times$16 = 10 + 120 + 126, where only
the 126 contains an $SU(5)$ singlet component.  The most
straightforward possibility for generating a term like
Eq. (\ref{eq:NMassTerm}) in the full theory is therefore to include a
Higgs 126, and a Yukawa coupling of this to the 16s.  If the
appropriate components of the 126 acquire vacuum expectation values,
Eq. (\ref{eq:NMassTerm}) will emerge.  The 126 is a five-index
self-dual antisymmetric tensor under $SO(10)$, which may not be to
everyone's taste.  Alternatively, one can imagine that more
complicated interactions, containing products of several simpler Higgs
fields which condense, are responsible.  These need not be fundamental
interactions (they are, of course, non-renormalizable), but could
arise through loop effects even in a renormalizable field theory.

At this level there are certainly many more options than constraints,
so that without putting the discussion of N masses in a broader
context, and making some guesses, one can't very specific or
quantitatively precise.  Nevertheless, I think it is fair to say that
these general considerations strongly suggest that $\eta$ is
associated with breaking of unified symmetries down to the Standard
Model.  Thus, if the general framework is correct, the expected scale
for its entries is set by the one we met in the unification of
couplings calculation, i.e. $\eta \sim$10$^{16}$ Gev.

The Ns communicate with the familiar fermions through the Yukawa interactions 

\begin{equation}
\Delta {\cal L}_{N-L} ~=~ g^i_j {\bar N}_i L^{aj} \phi^\dagger_a~+~{\rm h.c.}~,
\label{eq:NLCoupling}
\end{equation}
using the previous notations but now, in this `conventional' term,
suppressing the Dirac spinor indices.  These interactions are of
precisely the type that generate masses for the quarks and charged
leptons in the Standard Model.  If N were otherwise massless, the
effect of Eq. (\ref{eq:NLCoupling}) would be to generate neutrino
masses, of the same order as ordinary quark and lepton masses.  In
$SO(10)$, indeed, these masses would be related by simple
Clebsch-Gordon and renormalization factors of order unity.
Fortunately, as we have seen, N is far from massless.


Indeed, it is so massive that for purposes of low-energy physics we
can and should integrate it out.  This is easy to do.  The effect of
combining Eq. (\ref{eq:NMassTerm}) and Eq. (\ref{eq:NLCoupling}) and
integrating out N is to generate
\begin{equation}
\Delta {\cal L}_{\rm eff.} ~=~ g^k_i g^l_j (\eta^{-1})_{kl} L^{\alpha
a i} L^{\beta b j} \epsilon_{\alpha \beta} \phi^\dagger_a
\phi^\dagger_b + {\rm h. c.} ~.  
\label{eq:Leff}
\end{equation}
Thus we arrive back at Eq. (\ref{eq:newL}), with 
\begin{equation}
\kappa_{ij} ~=~ g^k_i g^l_j (\eta^{-1})_{kl}~. 
\label{eq:seesaw}
\end{equation}

This ``seesaw'' equation provides a much more precise version of the
loose connection between unification scale and neutrino mass we
discussed at the outset.  There is much uncertainty in the details,
since there is no reliable detailed theory for the $g^k_i$ nor the
$\eta$s.  But if $g$ has an eigenvalue of order unity pointing toward
the third family (as suggested by symmetry and the value of the top
quark mass), and if we set the scale for $\eta$ using the logic above,
then we get close to $10^{-2}$ eV for the $\tau$ neutrino mass, as
observed.

While at present it is less imposing than the others, this success
promises to become the third pillar of unification.

The pattern of quark and charged lepton masses suggests that the other
eigenvalues of $g$ might be considerably smaller, thus generating a
hierarchical pattern of neutrino masses.  This is at least broadly
consistent with proposed explanations of the solar neutrino anomalies,
but will not readily accommodate the reported LSND results, nor
neutrinos as cosmologically significant hot dark matter.

\section{Summary and Prospect}

A mass of approximately 10$^{-2}$ eV for the heaviest neutrino fits
beautifully into the framework of supersymmetric unification in
SO(10).  This sort of theory unifies the fermions in a particularly
compelling way, with all the quarks and leptons in a generation
fitting into a single multiplet, but requires the existence of new
degrees of freedom, the Ns (one per family), which within the theory
are predicted to be very heavy.  The Ns themselves are not accessible,
but they induce tiny masses for the observable neutrinos.  Assuming
supersymmetry is spontaneously and only mildly broken, this sort of
theory also has impressive quantitative success in accounting for the
disparate values of the gauge couplings of the Standard Model.
Although I don't have time to discuss it here, one also finds here
an attractive mechanism for understanding why the standard model Higgs
field, unlike the other ingredients of the Standard Model, forms an
incomplete multiplet of the unified symmetry \cite{dimo}.

In this talk I have taken a minimalist approach, extrapolating
straight weak-coupling quantum field theory and gauge symmetry up to
near- (but sub-) Planckian mass scales, using only degrees of freedom
that the facts more or less directly require.  This approach has the
advantage of allowing us to make some simple, definite predictions.

General consequences of the minimalist framework are that the neutrino
masses are Majorana and that there are no light sterile neutrinos.
Also, it is hard to avoid a hierarchical pattern of neutrino masses.
This makes it difficult to accommodate a cosmologically significant
contribution of neutrino dark matter.  These are eminently falsifiable
assertions.  Indeed, at this conference some have argued, implicitly
or explicitly, that they already have been falsified.  We shall see.
If the minimalist framework really does break down, we will have
learned a profound lesson.

The large mixing angle indicated by the atmospheric oscillation
results, though by no means problematic, does come as something of a
surprise.  To do justice to experimental information at this level of
detail, we must consider it in conjunction with the whole complex of
questions around how unified symmetry is broken and how the pattern of
quark and lepton masses is set.  Some general considerations that
guide this sort of phenomenology were discussed here by Professor
Pati, and in rather different ways by Professors Langacker, Mohapatra,
Ramond and Yanagida.  In working on this subject with Babu and Pati, I have
been pleasantly surprised at how well so many diverse facts can be fit
together.  But as yet no insight comparable to the ``pillars''  has
emerged from thinking about the pattern of masses and mixings, and
here one longs for a deeper, more compelling theory.

In any case, the acid test for this whole line of development is
nucleon instability.  Supersymmetric unification introduces new
sources of nucleon instability that are precariously close to existing
experimental limits.  The large mixing indicated by the atmospheric
neutrino oscillation results sharpens the problem from Higgsino
exchange, because the dangerous Higgsino exchange is suppressed by the
supposed smallness of its couplings to the light particles, and the
straightforward relation of mass to coupling will be modified by
mixing.  Also, careful inclusion of the fields necessary to break the
unified symmetry and generate neutrino masses brings to light
additional potential sources of nucleon instability \cite{bpw}.

I hope and expect that at some future conference we will hear from
SuperK -- or their successors -- reports of the other shoe dropping.


\end{document}